# Phosphorene quantum dot electronic properties and gas sensing

Hazem Abdelsalam[1*], Vasil A. Saroka[2], Waleed O. Younis[3,4]

[1]Theoretical Physics Department, National Research Centre, El-Buhouth Str., Giza, 12622, Egypt
[2]Institute for Nuclear Problems, Belarusian State University, Bobruiskaya 11, 220030 Minsk, Belarus
[3]Physics Department, Faculty of Science, Beni-Suef University, Beni-Suef City, 62521, Egypt
[4] Vice Rectorate for Graduate Studies and Scientific Research, Imam Abdulrahman Bin Faisal University, Dammam, 1982, Saudi Arabia

## ABSTRACT

Density functional theory calculations are performed on phosphorene quantum dots having different shapes and edge terminations to investigate their structure stability, electronic properties, and gas sensing ability. All the selected phosphorene dots, namely hexagonal and triangular flakes with armchair and zigzag terminations, have positive binding energies which insure their stability even though the bond lengths are much longer than those in the infinite phosphorene layer. It is found that all the selected hydrogen passivated quantum dots have a wide energy gap. In contrast, the partial passivation with sulfur decreases the gap. Moreover, it transforms the system from antiferromagnetic to ferromagnetic state. The energy gap of hexagonal zigzag cluster can be additionally tuned by electric field: narrowed by about 1.7 eV for hydrogenated or broadened by 0.25 eV for partially sulfurated edges. It is shown that phosphorene quantum dots successfully adsorb $H_2S$, $CH_4$, $CO$, $NH_3$ gas molecules either on their edge or surface. The highest adsorption energy is obtained for $NH_3$ molecule, when it is placed over the surface. This adsorption is alleviated by in-plane electric field and hindered by perpendicular field.

## 1. INTRODUCTION

Mainstreaming nanotechnology two dimensional materials provide a unique and versatile playground for the theoretical and experimental physics. A number of such materials have been synthesized in past few decades: graphene, silicene, phosphorene etc. [1-2]. These materials have attractive properties for the next generation electronics and optoelectronics [3]. Moreover, their large surface area and ultra-small thickness also provides a great sensitivity to the biological, organic and gas molecules [4]. One of the novel 2D structures that exist in a free standing form is phosphorene – a monolayer of black phosphorous [5-7]. In contrast to graphene it has a puckered lattice structure and strong intralayer anisotropy of the electronic and optical properties [8]. Recently, black phosphorous quantum dots have been synthesized by liquid exfoliation technique [9,10], which has fostered an intensive theoretical investigation of singly and a few-layer phosphorene quantum dots within the tight-binding models [11-15]. In contrast to graphene [16] or silicene [17] quantum dots, in the tight-binding model the majority of phosphorene quantum dots possess the so-called quasi-zero energy states. These states are localized within the energy gap independently of quantum dot edge morphology and are sensitive to the external perturbations of electrostatic nature [12]. Such perturbations can be imposed by an external electric field or adsorption of the gas molecules on edges or surface of a quantum dot. In latter case an enhanced sensitivity of such physical quantities as electrical conductivity or optical absorption to the adsorbed molecules may be expected. The sensing capability of graphene and silicene with respect to such important gas molecules as NH3, NO, CO has been paid attention [18-21]. Some comprehensive reviews can also be found [22]. Recently, it has been also demonstrated experimentally that studied theoretically for some time nanoribbons with complex shapes of edges, such as zigzag-shaped ribbons [23-26], exhibit enhanced sensitivity to low molecular weight alcohols [27]. Thereby the decisive role of edges in sensing applications has been revealed. At the same time, although detection schemes have been proposed for phosphorene [28], the potential of phosphorene structures with edges, such as quantum dots, stays largely unexplored with respect to gas sensing [29].

In this paper we investigate the electronic properties of edge functionalized phosphorene quantum dots (PQDs) subjected to electric field using density functional theory (DFT) [30,31]. The selected PQDs have hexagonal and triangular shapes with armchair and zigzag terminations. The abbreviation AHEX refers to armchair hexagonal, ATRI refers to armchair triangular. ZTRI and ZHEX refer to zigzag hexagonal and triangular flakes, respectively. All the DFT calculations were implemented with Gaussian 16 program [32]. The hybrid exchange-correlation functionalB3LYP (Becke, three-parameter, Lee-Yang-Parr) [33,34] and the 3-21g basis set was employed in the optimization process. It was found that the 3-21g basis is sufficient basis set for DFT calculations in graphene and silicene quantum dots [35,36], when considering the computational power and the results accuracy. Therefore, it is adopted here as a sufficient basis set to describe the phosphorus atoms in the phosphorene clusters.

## 2. RESULTS AND DISCUSSION

Controlling the electronic properties of PQDs by edge modification and electric field makes them potential applicant for various devises such as gas sensors. Here we study the effect of partial sulfuration and electric field on the electronic properties of edge hydrogenated PQDs. Moreover, the sensing capability of PQDs to detect such gases as H2S, CH4, CO, and NH3 will be examined.

### 2.1. STRUCTURAL PROPERTIES

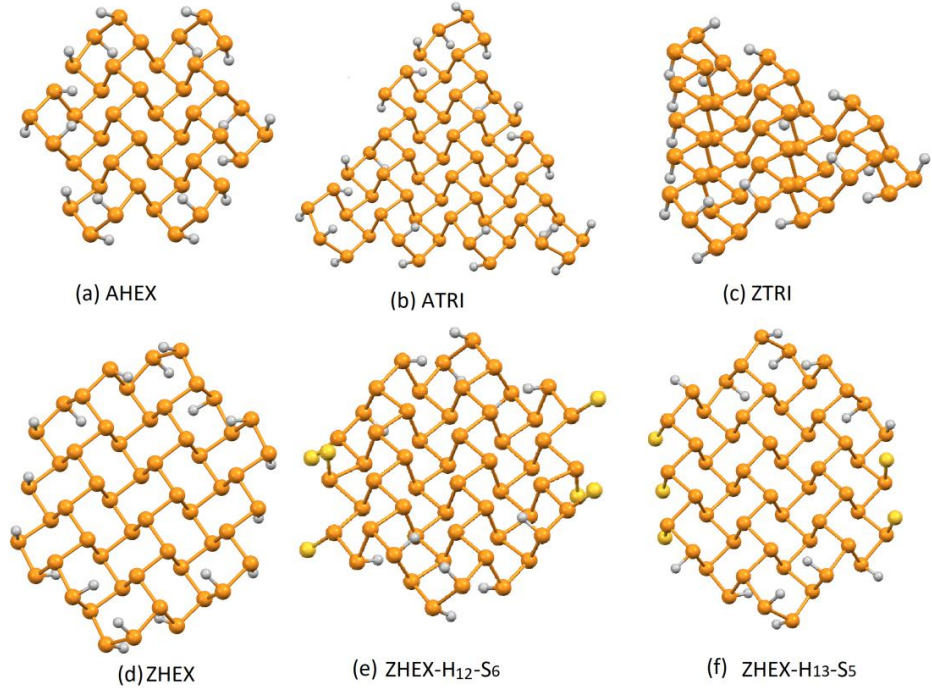

Fig. 1. The equilibrium structures of hexagonal and triangular PQDs saturated with H (a), (b), (c), and (d). ZHEX PQD with partial sulfuration (e), (f).

The optimized structures of hexagonal and triangular phosphorene dots with armchair and zigzag terminations are shown in Fig. 1. The edges are saturated with H as seen in Fig. 1 (a), (b), (c), and (d). In order to study the interaction of edge atoms with other elements (such as S) and to reveal its effect on the structure stability and electronic properties, we consider ZHEX partial passivation with S as shown in Fig. 1 (e) and (f). The bond length between the H and edge P atoms are the same ($d_{HP}$=1.44-1.45Å) in all the selected flakes, see Table 1. In ZHEX dot the bond lengths between S and edge P atoms $d_{SP}$=1.44-2.29Å is higher than $d_{HP}$. The bond lengths between phosphorus atoms ($d_{P-P}$) range from 2.37 Å to 2.49 Å within the cluster. This is 9.5-12.8 % higher than the bond lengths in infinite phosphorene layer. Recall that the distances between two P atoms within the sublayer and in different sublayers are 2.164 Å and 2.207 Å, respectively (see Fig.3 in ref. [12]). In order to verify the stability of the selected PQDs, we calculated their binding energies by using the following equation: $E_B= (N_H E_H + N_S E_S + N_P E_P - E_C)/N$. Where $E_B$ is the binding energy, $N_H$, $N_S$, $N_P$, and $N$ are the number of hydrogen, sulfur, phosphorus, and total number of atoms, respectively, the corresponding energies are $E_H$, $E_S$, $E_P$, and $E_C$, respectively. The values of $E_B$ for different clusters are shown in Table 1; from the positive values of the binding energy we infer that all the selected clusters with different edge passivation are stable. It is worth noticing that binding energies of the hydrogenated and partially sulfurated clusters are very close, $E_B \sim 3.7$-$3.8$ eV. With large number of S atoms attached to the edges, such as fully sulfurated edges, the interaction of sulfur second electron in p-orbital with P atoms at the edges leads to strong deformation of the cluster shape which makes this cluster being unstable. That is why we consider the partial passivation with S atoms presented in Fig. 1 (e).

Table 1. $d_{XP}$ is the distance between H or S atom and the edge P atom, $d_{P-P}$ gives the bond length between P atoms, $E_B$ is the binding energy, and $E_g$ is the energy gap.

| Structure | $d_{XP}$ (Å) | $d_{P-P}$ (Å) | $E_B$ (eV) | $E_g$ (eV) |
|---|---|---|---|---|
| AHEX ($P_{42}H_{18}$) | 1.44-1.45 | 2.37-2.43 | 3.65 | 2.96 |
| ATRI ($P_{60}H_{24}$) | 1.44-1.45 | 2.37-2.44 | 3.68 | 2.82 |
| ZTRI ($P_{46}H_{18}$) | 1.44-1.45 | 2.38-2.42 | 3.69 | 3.00 |
| ZHEX ($P_{54}H_{18}$) | 1.44-1.45 | 2.38-2.43 | 3.75 | 2.77 |
| ZHEX ($P_{54}H_{12}$-$S_6$) | 1.44-2.15 | 2.38-2.49 | 3.77 | 0.27 |
| ZHEX ($P_{54}H_{13}$-$S_5$) | 1.44-2.29 | 2.37-2.45 | 3.79 | 1.26 |

## 2.2. ELECTRONIC PROPERTIES

The electronic density of states (DOS) is introduced here to study the energy levels and energy gap of the PQDs under the effect of shape (hexagonal and triangular), edge termination (armchair and zigzag), and chemical modification.

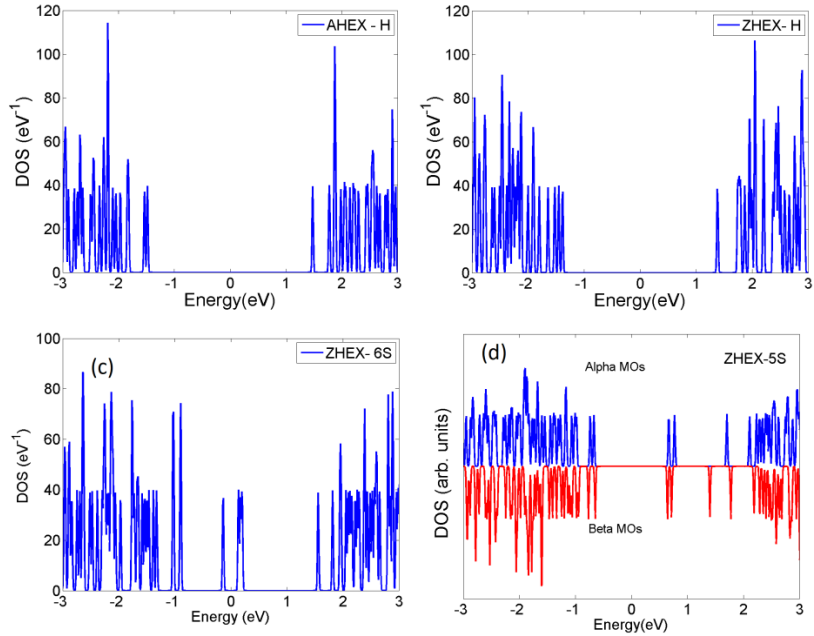

Fig. 2. The electronic density of states of different PQDs: (a) fully hydrogenated ZHEX, (b) fully hydrogenated AHEX, (c) and (d) partially sulfurated ZHEX.

Each energy levels is represented by Gaussian function $\frac{1}{\sqrt{2\pi}\alpha}\exp\left[\frac{-(\varepsilon-\varepsilon_i)^2}{2\alpha^2}\right]$ with broadening $\alpha = 0.01\ eV$.

The Fermi level is hold fixed, $E_F = (E_{HOMO}+E_{LUMO})/2$, with $E_{HOMO}$ denotes the energy of the highest occupied molecular orbital (HOMO) and $E_{LUMO}$ the energy of the lowest unoccupied molecular orbital (LUMO). Fig. 2 shows the DOS of AHEX, ATRI, ZTRI, and ZHEX quantum dots. As can be observed from Fig. 2 and from the values of $E_B$ in Table 1, the energy gap in all the hydrogenated phosphorene flakes are almost the same which means that all the selected flakes are insulator regardless of the shape or the edge termination. This energy gap decreases by increasing the cluster size, for instance for AHEX ($P_{222}H_{42}$) the energy gap becomes 2.03 eV. In addition to cluster size, the chemical modification can be used to tune the energy gap. Edge relaxation by S atoms provides a transformation from insulator PQDs as in ZHEX-H ($E_g$=2.77eV) to a narrow gap semiconductor PQDs as in ZHEX-$S_6$ ($E_g$=0.27 eV). This transformation occurs due to the unbound electrons from the sp3 hybridized S atoms, these electrons have low binding energy (highly reactive) and therefore fill the energy gap. Interestingly, the partial passivation with sulfur can change the magnetic state of the phosphorene flake depending on the number of attached S atoms. The ZHEX ($P_{54}H_{18}$) experiences a transformation from the antiferromagnetic to ferromagnetic state when 5 H atoms are replaced by 5 S atoms, as seen in Fig. 2 (d) for ZHEX ($P_{54}H_{13}$-$S_5$). The energy gap ($E_g$=1.26 eV is the energy difference between the LUMO in the beta molecular orbitals and the HOMO in the alpha orbitals energies) in this case is intermediate between that of ZHEX ($P_{54}H_{18}$) and ZHEX ($P_{54}H_{13}$-$S_6$). Thus, we see that chemical functionalization provides flexible control of the electronic and magnetic properties of phosphorene quantum dots.

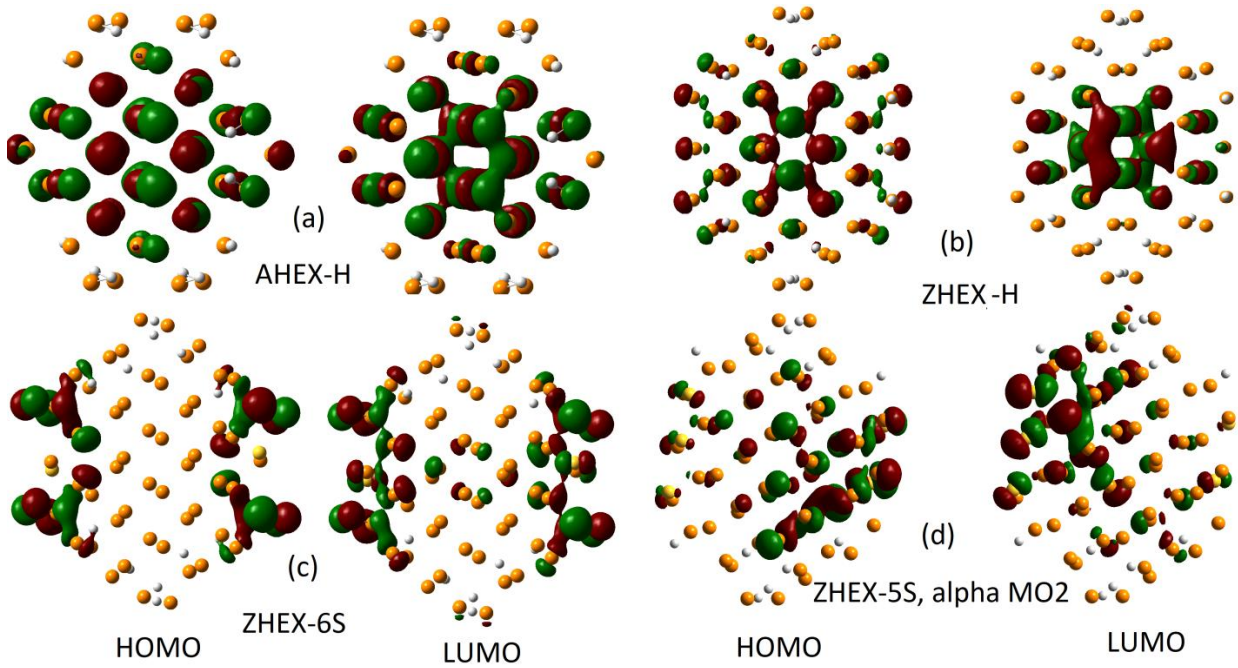

Fig. 3. The HOMO/LUMO for some selected PQDs under full hydrogenation (a, b) and partial sulfuration (c, d).

Fig. 3 presents the HOMO and LUMO distribution for some selected clusters, namely AHEX-H, ZHEX-H, ZHEX-6S, and ZHEX-5S. As seen from Fig. 3 (a) and (b) they are distributed over the cluster surface and are almost situated on single atoms which imply that they are weakly bound to the cluster. The origin of these states is the lone pair of electrons in the sp3 hybridized P atoms which do not interact with H or neighboring P atoms. On the other hand, when PQDs are partially passivated with S atoms, the distribution of the HOMO and LUMO is totally different. In case of ZHEX-6S, as shown in Fig. 3 (c), the HOMO and LUMO are localized on the edges. This distribution occurs due to the fact that the single unbounded electrons from S atoms at the edges are the ones that form the HOMO and LUMO. Therefore the ZHEX-6S is a semiconductor which is highly interactive only through its edges, while AHEX-H and ZHEX-H are insulators which interact with the surrounding through the cluster surface. The distribution of HOMO/LUMO states in the ZHEX-5S (Fig. 3 (d)) is intermediate between the fully hydrogenated ZHEX and the ZHEX-6S where they distribute over the surface and the sulfur atoms.

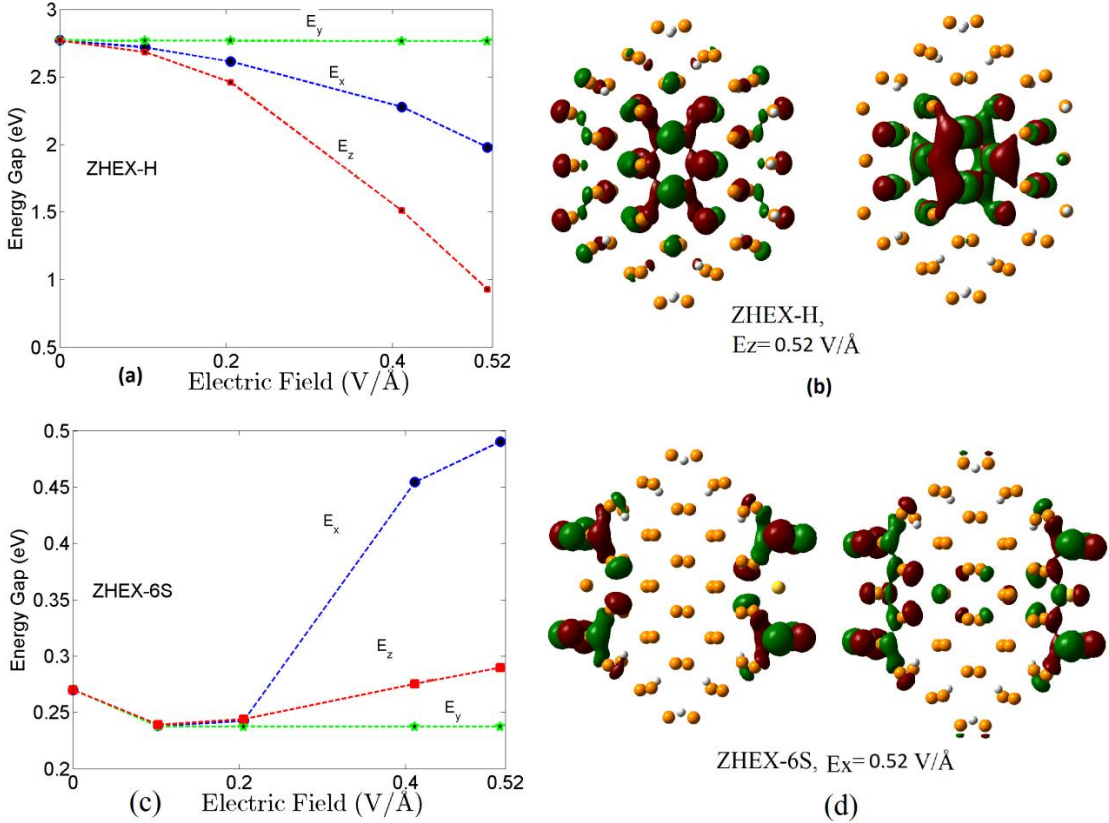

Fig. 4 The effect of electric field on the energy gap of ZHEX-H (a) and ZHEX-6S (c). The HOMO/LUMO distribution of ZHEX-H (b) and ZHEX-6S (d) at Ez= 0.52 V/ Å .

Having explored the electronic properties of PQDs under the effect of shape, edge termination, and chemical functionalization, we turn on to study the effect of electric field. The effect of electric field applied in x-, y- and z-directions ($E_x$, $E_y$ and $E_z$, respectively) is explored. We focus on the effect of the field on the energy gap and the HOMO/LUMO distribution. The ZHEX-H and ZHEX-6S have been chosen for studying this effect in PQDs. The former has a large energy gap, while the later represents the cluster with a small energy gap. Fig. 4 shows the variation of the energy gap of ZHEX-H (a) and ZHEX-6S (c) with electric field. In order to visualize the effect of electric field on molecular orbitals, we plot the HOMO/LUMO distribution in Fig. 4 (b) and (d) for fully hydrogenated and partially sulfurated ZHEX, respectively. As one can see in Fig. 4 (a), the electric field closes the large energy gap of the hydrogenated clusters: the perpendicular field ($E_z$) generates the largest decrease in the energy gap while the effect of the in plane $E_y$ field is nearly negligible. The already small energy gap as in ZHEX-6s (Fig. 5 b) decreases with increasing the electric field until E= 0.2 V/Å, then it starts to increase reaching a value of $E_g$=0.5 eV at $E_x$=0.52 V/ Å. The HOMO/LUMO distribution is almost unaffected by the applied field, which means that the interactive electrons and their sites on the edges or over the whole cluster are unaffected by the electric field.

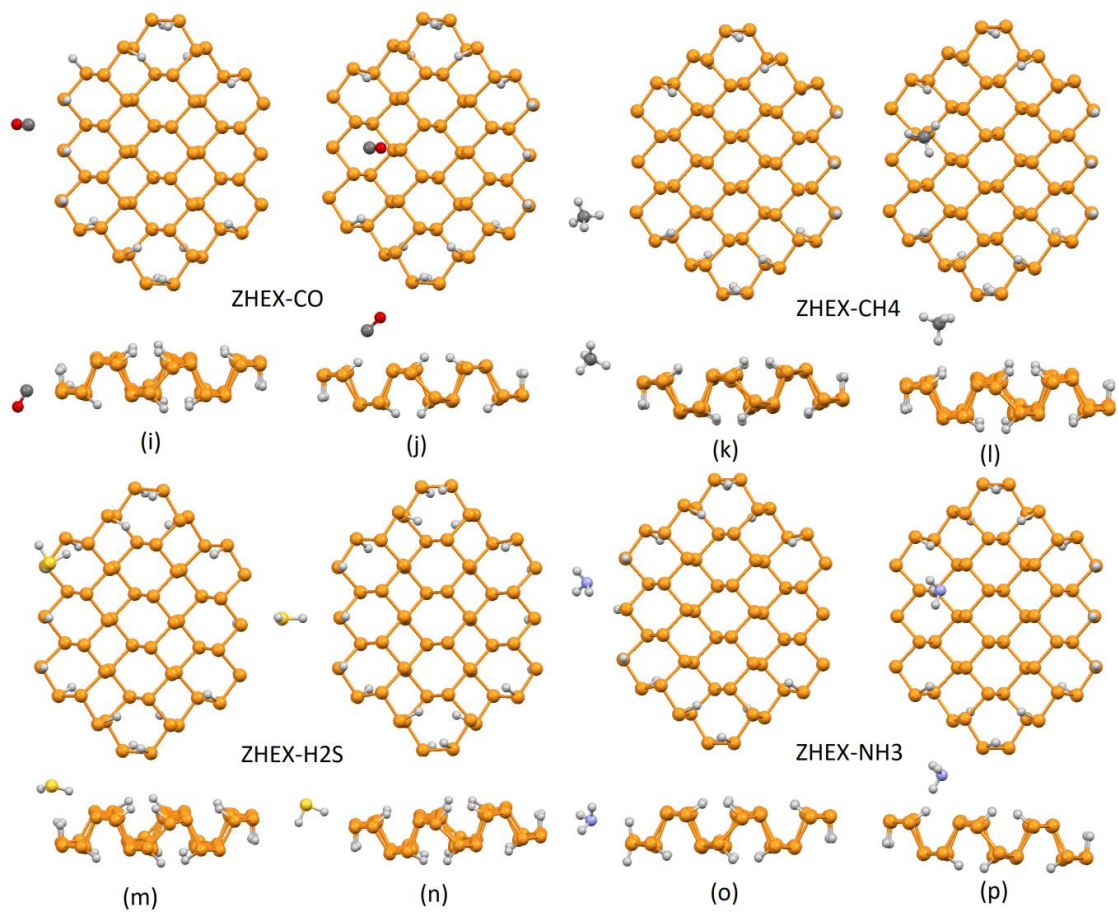

Fig. 5. Adsorption of different gas molecules on the edge and above the surface of hydrogenated ZHEX.

## 2.3. SENSING CABAPILITY OF PQDs

In order to study the capability of PQDs as a potential gas sensor we plot Fig. 5 that shows the optimized structure of ZHEX after adsorption of different gases. The gases considered here are $H_2S$, $CH_4$, CO, and $NH_3$. The attachment of the gas to the edge or the surface of the phosphorene dot is initially considered as placing the gas at distance =3 Å from the edge or above the surface. Then DFT calculations are performed to estimate the stability and the adsorption position of the gas on the phosphorene quantum dot. The adsorption positions (D) is given in Table 2. The adsorption energy ($E_{ad}$) of the gas molecule on PQDs is obtained by the following equation: $E_{ad}= (E_{gas}+E_{PQDs}–E_{total})$, where $E_{gas}$, $E_{PQDs}$, $E_{total}$ denote the total energy of the gas molecule, the phosphorene dot, and the gas molecule + phosphorene quantum dot system, respectively. According to this definition, a positive value of $E_{ad}$ ensures the stability of the resultant structure. The larger the positive value of $E_{ad}$ the stronger adsorption of gas molecule on the PQD is. The calculated positive adsorption energies, given in Table 2, imply that all the selected gases favorably adsorb on the edge or surface of the PQDs. $NH_3$ has the highest adsorption energy while $CH_4$ has the lowest adsorption energy. All the gases can be adsorbed on the edge or the surface of the PQDs with approximately the same adsorption positions and energies except for the adsorption of $H_2S$ whose adsorption strength is noticeably higher at the edge than on the surface. The charge transfer ($\Delta Q = Q_{total} - Q_{PQDs} - Q_{gas}$, with $Q_{total}$ represents the system net charge after adsorption, $Q_{PQDs}$ the charge before adsorption, and $Q_{gas}$ is the gas charge) has also been calculated to reveal its influence on the adsorption energy. The charge transfer is calculated based on the Hirshfeld charge analysis [37]. We found that there is a strong correlation between the adsorption energy and the amount of charge transfer. For instance, the charge transfer from the ZHEX to CO molecule is $0.73 \times 10^{-3}$ (e) and to $NH_3$ is $2.6 \times 10^{-3}$ (e); the negative sign represent transfer of charge from the PQDs to the gas and the inverse transfer for positive sign. The corresponding adsorption energies are 0.17 and 0.69 eV, respectively, which implies that the increase in the magnitude of charge transfer is accompanied by increase in the adsorption energy. Moreover, it can be seen from Table 2 that with respect to each gas molecule the adsorption energy is higher in small adsorption positions than that with large adsorption positions. According to the above discussion we conclude that the energetically preferable adsorption position of $H_2S$, see Fig. 5(e), is the edge adsorption with $E_{ad}$=0.16 eV. The optimal adsorption positions and adsorption energy of all the four gases are highlighted in bold in Table 2 (for $CH_4$ edge and surface $E_{ad}$ equal 0.0373 and 0.0423 eV, respectively, therefore we highlight $E_{ad}$ on the surface). The values of $E_g$ given in Table 2 imply that the energy gap slightly decreases in the cases of strong absorption as noticed in the surface adsorption of $NH_3$.

Table 2. The adsorption distance (D) of gas molecules at the edge and above the surface of the ZHEX, the corresponding charge transfer ($\Delta Q$), adsorption energy ($E_{ad}$), and the energy gap ($E_g$).

| Structure | Gas | Adsorption site location | $D$(Å) | $\Delta Q x 10^{-3}$ (e) | $E_{ad}$ (eV) | $E_g$ (eV) |
|---|---|---|---|---|---|---|
| **ZHEX** ($P_{54}H_{18}$) | $H_2S$ | Edge | **3.68** | -0.16 | **0.16** | 2.76 |
| | | surface | 4.89 | 0.003 | 0.04 | 2.77 |
| | **CH₄** | edge | 3.91 | -1.7 | 0.04 | 2.77 |
| | | surface | **3.78** | -1.8 | **0.04** | 2.77 |
| | CO | edge | **3.12** | -0.73 | **0.17** | 2.77 |
| | | surface | 3.23 | -0.81 | 0.14 | 2.78 |
| | $NH_3$ | edge | 2.70 | -2.1 | 0.64 | 2.68 |
| | | surface | **2.52** | -2.6 | **0.69** | 2.66 |

Now we proceed to discussion of the dependence of the adsorption energy on the electric field. The adsorption of H2S (lowest $E_{ad}$) and NH3 (highest $E_{ad}$) on the surface of ZHEX PQD were chosen for a model study of PQD sensing capability in the electric field. The adsorption energies of NH3 (a) and H2S (b) on the surface of ZHEX-H are shown in Fig. 6. It is evident that the adsorption capability of gas molecules by phosphorene clusters can be considerably enhanced by application of in plane electric field ($E_x$ and $E_y$). On the contrary, a perpendicular field decreases that adsorption. The very small adsorption energy of H2S, Ead=0.04 eV, can be increased manyfold by in-plane electric field up to 0.25 eV at $E_x$ or $E_y$ ~ 0.52 V/Å. Hence the electric field allows one to tune the sensing capabilities of PQDs.

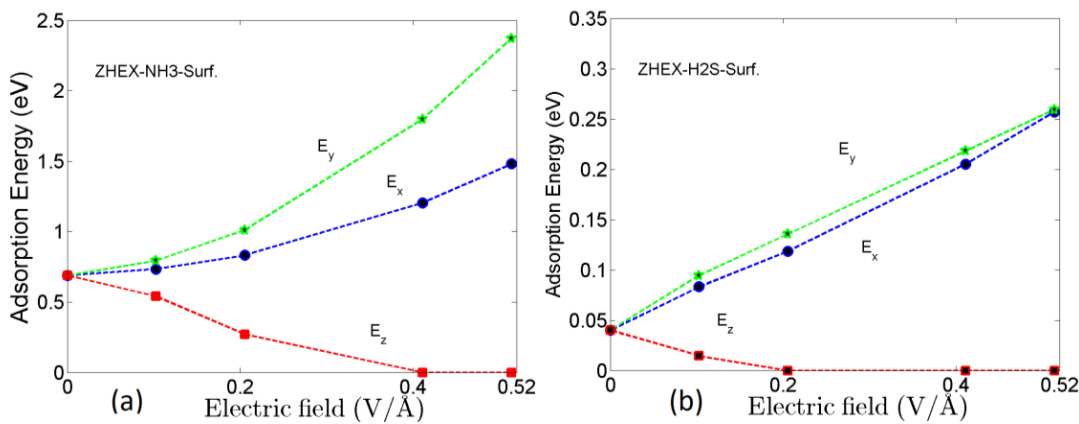

Fig. 6. The adsorption energy of $NH_3$ (a) and $H_2S$ (b) on the surface of ZHEX-H subjected to in-plane ($E_x$ and $E_y$) and out-of-plane ($E_z$) electric fields.

## 3. CONCLUSION

In this paper, we have investigated the electronic properties and the gas sensing capability of phosphorene quantum dots that have different shapes and edge terminations. Four types of phosphorene quantum dots have been selected: with hexagonal and triangular shapes and with armchair and zigzag terminations. The optimized structures are characterized by higher bond lengths with respect to the infinite single-layer phosphorene. However, all the clusters under hydrogenation and partial sulfuration are stable. The electronic properties can be precisely controlled, the energy gap can be decreased from 2.77 eV to 1.26 eV or even 0.27 eV, when the passivation of hexagonal-zigzag quantum dot is changed from fully hydrogenated to partially sulfurated with 5 or 6 S atoms. Additionally, the magnetic state can be transformed from antiferromagnetic state as in hydrogenated hexagonal-zigzag to ferromagnetic state with non-zero total spin as in partially sulfuration with 5 S atoms. The energy gaps are most efficiently tuned by applying external electric field perpendicular to the plane of the quantum dot (z-direction) and in plane of structure perpendicular to zigzag crystallographic orientation (x-direction). The energy gap in fully hydrogenated hexagonal zigzag decreases by increasing the electric field while it moderately increases with the applied field in the same clusters with partial sulfuration with six S atoms. We have also found that $H_2S$, $CH_4$, CO, $NH_3$ molecules can be adsorbed on the edge or the surface of the phosphorene quantum dots with zigzag edges and hexagonal structure. The maximum adsorption energy is observed for surface adsorption of NH3. Most importantly, the application of perpendicular electric field decreases the adsorption energy and can prevent this process in high electric field (0.2-0.4 V/ Å). Concurrently, the in-plane electric field ($E_x$ and $E_y$) provides up to 4 times enhancement of the adsorption energy at electric field of about 0.5 V/Å. Such tunability of adsorption should be useful in designing superior phosphorene gas sensors.

## ACKNOWLEDGMENTS

This research utilized Imam Abdulrahman Bin Faisal (IAU)'s Bridge HPC facility, supported by IAU Scientific and High Performance Computing Center. https://doi.org/10.5281/zenodo.1117442. VAS acknowledges funding by EU H2020 RISE project CoExAN (Grant No. H2020-644076).

## REFERANCES